# Spatio-temporal Modulation Instability of Surface Plasmon Polaritons in Graphene-dielectric Heterostructure

**Morteza A. Sharif[1]**

1. Optics and Laser engineering group, Faculty of electrical engineering, Urmia University of Technology, Band road, Urmia, Iran, Email address: m.abdolahisharif@ee.uut.ac.ir

**Abstract-** Using the Jacobi Elliptic Functions, an analytical solution is developed for the nonlinear amplitude equation of Surface Plasmon Polaritons (SPPs) in a graphene-dielectric waveguide. It is shown that the field localization of SPPs coupled with TM polarized terahertz light can be enhanced if the nonlinearity is increased. On the side, a numerical solution based on Split Step Beam Propagation Method (SSBPM) suggests that spatial Modulation Instabilty (MI) can be dominant. Accordingly, larger nonlinearity leads to the generation of discrete plasmon solitons rather than the diffracted profile resulted for the modest nonlinearity. Adding then the temporal variations to the nonlinear amplitude equation and solving numerically by predictor –corrector method, it is revealed that temporal MI appears as ultrashort pulse trains with multi-periodic behavior. Evoking the similarity with a laser cavity, the waveguide can be assumed as a spasing system - assuming the large nonlinearity regime- in which if the coupling depth is raised, the character of MI will be changed from the convective to absolute and the amplitude of SPPs will grow fast. The procedure can be even chaotic and unpredictable. This spasing system is suitable for applications of the electro-optical circuitry, optical amplification, optical communication and biomedical sensing through which the low power consumption and non-destructivity are of important traits.

## 1- Introduction

Strong nonlinear response of graphene- the one-atom thick material with hexagonal lattice- to the external electromagnetic wave has been confirmed by several experimental and theoretical studies[1-8]. Tunability and ultrafast dynamics are the two other nonlinear characteristic features of graphene[9-10]. The nonlinear response can be extremely enhanced if a graphene-dielectric heterostructure is used[11-18]. Based upon, the propagation of optical solitary waves can be realized. Maxi L. Nesterov et al have shown that the optical solitons can be supported within a single layer graphene sandwiched between the two dielectric slabs[19]. As well, Yu V. Bludov, et al have demonstrated the presence of discrete solitons for multiple graphene sheets separated by dielectric layers[20].

During the preparation stage of graphene, lots of graphene Quantum Dots (QDs) may be formed within a sheet[21]. If irradiated with a strong cw pump wave, these QDs will act as nano-optical pulsers among which the energy can be transferred by near field interactions[22]. Subwavelength light-substantial particles interfering in these interactions are then called Surface Plasmon-Polaritons (SPPs). Overcoming the diffraction limit of light, SPPs can carry a large field localized within a tiny propagation length [23-28]. Accordingly, novel graphene based SPPs' waveguides have been proposed to be applicable in the fields of optical communication and biomedicine[29-34]. Meanwhile, researchers have intending to investigate the effect of nonlinearity on the robustness of plasmonic waves. In this connection, A. V. Gorbach has derived an amplitude nonlinear wave equation for surface plasmonic waves assuming a suspended monolayer graphene in a dielectric medium. He has investigated the effect of nonlinearity on TM and TE polarized plasmonic waves[35]. In corroboration, Sergey Mikhailov has studied the effect of strong pump wave on the wavelength and propagation length of plasma resonances in graphene[36]. Already, Yu. V. Bludov, et al have distinguished the subwavelength plasmon/photon solitons [20].

On the other hand, spatio/temporal Modulation Instability (MI; also known as Benjamin-Feir instability) appears through the different wave phenomena in consequence of interplay between the nonlinearity and diffraction/dispersion [37-40]. Hereof, plasmonic waves are not excluded. D. A. Korobko et al have investigated the evolution of SPPs induced by MI in a thin metal film. They have shown that the weak perturbations due to MI can grow fast such that the SPPs are transformed to the pulse trains[41]. S. G. Moiseev et al have also indicated the dependence of MI gain on the nonlinearity in a dielectric–metal–dielectric structure [42]. Despite all that, the process of MI has not been contemplated in a graphene-based plasmonic waveguide neither for spatial nor for its temporal feature. Furthermore, any investigation will also need considerations about unstable dynamical behavior.

In this study, a graphene-dielectric heterostructure is assumed as a plasmonic waveguide. Starting with an analytical solution based on the Duffing oscillator model for the nonlinear plasmonic amplitude equation, the effect of nonlinearity on plasmon soliton width is investigated. Considering the condition in which the balance between the nonlinear response and dispersion/diffraction can be violated, a numerical solution of nonlinear wave equation is then presented in order to obtain the spatial figure of SPPs subjected to MI. As the next step, the time variations of SPPs are added to the nonlinear amplitude equation. Consecutively, the temporal behavior of MI is investigated influenced by the nonlinearity and coupling depth.

## 2- Theoretical approach

The surface conductivity of single layer graphene can be obtained by Eq.(1) whereas Eq. (2) will give the simplified relation for intraband third order nonlinear conductivity if the third harmonic generation is assumed as the governing nonlinear phenomena[7].

$$\sigma^{(1)}(\omega) = i\left(\frac{e^2}{4\pi\hbar}\ln\left|\frac{2\mu-(\omega+i\tau^{-1})\hbar}{2\mu+(\omega+i\tau^{-1})\hbar}\right| + \frac{e^2 k_B T}{\pi\hbar^2(\omega+i/\tau)}\left[\frac{\mu}{k_B T}+2\ln(e^{-\frac{\mu}{k_B T}}+1)\right]\right), \quad (1)$$

$$\sigma^{(3)} = -\frac{9}{8}i\left(\frac{e^2}{\pi\hbar^2}\right)\left(\frac{(ev_F)^2}{\mu\omega^3}\right)(1+i\alpha_T), \quad (2)$$

$e$ is the electron charge, $\hbar$ is the reduced Planck's constant, $\omega$ is the angular frequency of the optical wave, $\mu$ is the chemical potential of graphene which is nearly equal to the Fermi energy $E_F$ assuming the low temperature regime[7]. $k_B$ is the Boltzmann's constant, $T$ is the temperature, $v_F$ is Fermi velocity and $\alpha_T$ is the nonlinear absorption coefficient.

A graphene-dielectric heterostructure is assumed as shown in Fig. 1. Single layer graphene is embedded inside two dielectric slabs with the dielectric constants $\varepsilon_1, \varepsilon_2$ respectively. The whole structure is supplied with double-gated electrode system. Top and back gate is used to apply an external perpendicular electric field which will modulate the Fermi energy level and subsequently the nonlinear optical conductivity (Eq.(2)). The Sided Source-Drain electrode system is also used to construct an ambipolar conductive structure in order to increase the mobility of charge carriers. This in turn has considerable influence on the generation rate of SPPs [43-44]. Generally, the Fermi energy is depended on the required gate voltage as given in Eq.(3).

$$E_F = \hbar v_F \sqrt{\pi \left(n_0 + C|V_g|/q\right)}, \qquad (3)$$

where $n_0$ is the intrinsic charier density; $C$ is the effective capacitance and $V_g$ is the gate voltage. Eq. (3) can be re-written as $E_F = \hbar v_F (\pi n_g)^{1/2}$ where $n_g$ is the total carrier density including the carriers due to the gate-induced doping[45].

The propagation of SPPs are assumed to be in $z$ direction while the transversal variations of the electric field will appear in unbound $y$ direction. Otherwise, there will be no principal difference between $y$ and $z$ direction due to the two dimensional nature of SPPs. External illumination with terahertz frequency is coupled with the waveguide. TM polarized SPPs can be excited as a result of modulating the nonlinear conductivity. A detailed explanation of the modulator working mechanism and SPPs' generation procedure can be found in other studies[29,44,46-51]. According to Eq.(1) and Eq. (2), the first and third order susceptibility of graphene sheets can be written as Eq.(4) and Eq.(5) respectively[7,35].

$$\chi_{gr}^{(1)} = \sigma^{(1)} / \left(-i\omega\varepsilon_0 d_{gr}\right) \delta(x), \qquad (4)$$

$$\chi_{gr}^{(3)} \simeq \frac{\sigma^{(3)}}{\left(-i\omega\varepsilon_0 d_{gr}\right)} \delta(x), \qquad (5)$$

in which, $d_{gr}$ is the thickness of single layer graphene and $\delta$ is the Dirac function. Therefore, the nonlinear refractive index $n_2$ of single layer graphene can be estimated as given in Eq.(6)[7].

$$n_2 = 3\chi_{gr}^{(3)} / \left[4\varepsilon_0 c \left(1 + \chi_{gr}^{(1)}\right)\right], \qquad (6)$$

where $c$ is the free space light velocity.

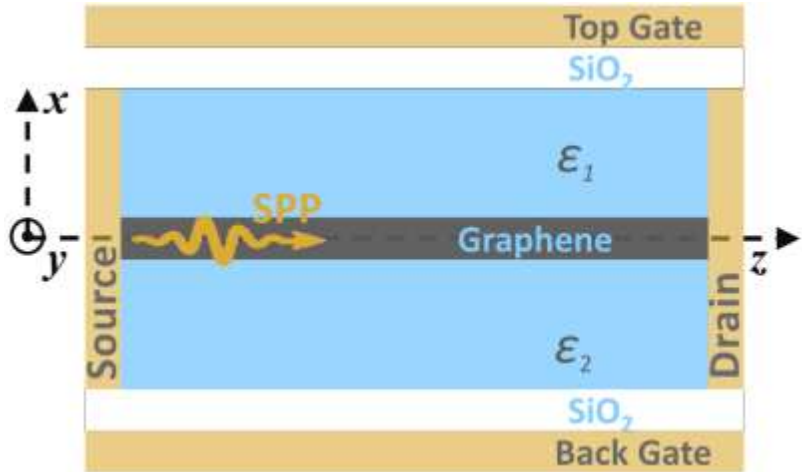

Fig.1. Schematic illustration of graphene-based SPP waveguide

*A- Plasmonic amplitude equation*

The SPPs' spatial nonlinear amplitude equation can be written as given in Eq.(7)[35].

$$2i\frac{\partial A}{\partial(z/k)}+\frac{1}{\beta k}\frac{\partial^2 A}{\partial(y/k)^2}+\gamma|A|^2 A+i\eta A=0, \quad (7)$$

$A$ is the slowly varying amplitude defined by $\vec{A}=I^{1/2}A(y,z)\vec{e}$ in which $I$ is the normalizing factor and $\vec{e}$ is the plasmon mode ; $k=2\pi/\lambda$ is the wave number in free space and $\lambda$ is the far infrared wavelength; $\beta$ is the propagation constant and is assumed dimensionless. In accordance, $|A|^2$ measures the power density per unit transversal length i.e. $y$. The coefficient $\gamma = g(\gamma_G + \gamma_D)$ is the total nonlinearity resulting from the contribution of single layer graphene and each dielectric slab nonlinearity; $\gamma_G$ and $\gamma_D$ can be calculated in terms of graphene nonlinear susceptibility and plasmon modes as indicated in Eqs.(8)-(12)[35].

$$\gamma_G = \frac{i\sigma_3 k^2}{2(\varepsilon_0 c\beta P)^2}\Theta\left[|\vec{e}|^2 |e_z|^2 + \tfrac{1}{2}\vec{e}^2(e_z^*)^2\right], \quad (8)$$

$$\gamma_D = \frac{k^2}{2\varepsilon_0^2 c(\beta P)^2}\int_{-\infty}^{+\infty}\chi_3\left(|\vec{e}|^4 + \tfrac{1}{2}|\vec{e}^2|^2\right)dx, \quad (9)$$

$$P = \int_{-\infty}^{+\infty}|\vec{e}|^2\,dx, \quad (10)$$

$$g = \frac{1}{(1+\xi)^2}, \quad (11)$$

$$\xi = \frac{-i}{\beta P}\Lambda[e_z^* e_x], \quad (12)$$

The operators $\Theta$ and $\Lambda$ are simply defined by the Dirac function as given in Eq.(13)[35].

$$\begin{cases}\Lambda(f(x))=\lim_{\delta\to 0}(f(x-\delta)-f(x+\delta)), \\ \Theta(f(x))=\tfrac{1}{2}\lim_{\delta\to 0}(f(x-\delta)+f(x+\delta)).\end{cases} \quad (13)$$

The parameter $\eta$ measures the linear absorption of graphene sheet and depends on the imaginary part of $\chi_{gr}^{(1)}$ as given in Eq.(14).

$$\eta = g^{1/2}\frac{\text{imag}(\sigma_1)k}{2\varepsilon_0 c\beta P}\Theta(|e_z|^2). \quad (14)$$

The function imag($s$) determines the imaginary part of $s$. Plasmon modes in Eqs.(8)-(14) are given in Eq. (15)[35].

$$\begin{cases}x<0: \vec{e}=\dfrac{-i\beta}{\kappa_2}\exp(\kappa_2 x)\hat{\mathbf{i}}+\exp(\kappa_2 x)\hat{\mathbf{k}} \\ x>0: \vec{e}=\dfrac{-i\beta}{\kappa_1}\exp(-\kappa_1 x)\hat{\mathbf{i}}+\exp(-\kappa_1 x)\hat{\mathbf{k}}\end{cases}, \quad (15)$$

where 1 and 2 stand for two successive dielectric slabs and the $\kappa_n$ denotes the propagation constant at $n$th dielectric slab as given in Eq. (16)[35].

$$\begin{cases}\kappa_n^2 = \beta^2 - \varepsilon_n, \\ \dfrac{\varepsilon_1}{\kappa_2}+\dfrac{\varepsilon_2}{\kappa_2}+\text{imag}(\dfrac{\sigma^1}{\varepsilon_0 c})=0,\end{cases} \quad (16)$$

The nonlinearity effect on the transversal variations of the mode amplitude $A$ is to be explored in Eq.(7). Assuming that the amplitude variations is known for a specific propagation length $z_0$ (i.e. $\left.\frac{dA}{dz}\right|_{z=z_0} = A_{z_0}$), the nonlinear wave equation can be re-written as Eq.(17).

$$\frac{1}{\beta k}\frac{\partial^2 A}{\partial (y/k)^2} + \gamma\left(\frac{k}{\beta}\right)^2 |A|^2 A + i\eta A = 0, \qquad \left.\frac{dA}{dz}\right|_{z=z_0} = A_{z_0}, \qquad (17)$$

where it should be noted that all terms are real due to the absorptive nature of the parameter $\eta$. Analytical solutions based on the Jacobi Elliptic Functions (JEF) can be developed for Eq.(17)[31,53]. The three principal elliptic functions are defined in the forms of Eq.(18).

$$\begin{cases} \text{sn}(u,\upsilon) = \sin\phi \\ \text{cn}(u,\upsilon) = \cos\phi \\ \text{dn}(u,\upsilon) = \sqrt{1-\upsilon^2 \sin^2\phi} \end{cases}, \qquad (18)$$

where $u = F(\phi,\upsilon) = \int_0^\phi \frac{d\vartheta}{\sqrt{1-\upsilon^2 \sin^2\vartheta}}$, $0 < \upsilon^2 < 1$; $\upsilon$ is elliptic modulus and $\phi = F^{-1}(u,\upsilon)$ is the Jacobi amplitude. Through a convenient notation, $k' = \sqrt{1-m}$ is introduced as an ancillary modulus whereas $m$ is substituted with $\upsilon^2$. To simplify, Eq.(17) can be rewritten as given in Eq.(19).

$$\frac{\partial^2 A}{\partial y^2} + \gamma'|A|^2 A + \eta' A = 0, \qquad \left.\frac{dA}{dz}\right|_{z=z_0} = A_{z_0}, \qquad (19)$$

in which $\eta' = i\eta$. A solution based on JEF can be written as Eq.(20)[31,53].

$$A(y) = c_1 \text{cn}\left(\sqrt{\eta' + c_1^2 \gamma}\, y + c_2, \sqrt{\frac{c_1^2 \gamma}{2(\eta' + c_1^2 \gamma)}}\right), \qquad (20)$$

where $c_1$ and $c_2$ are constants obtained by initial boundary conditions that are given in Eq.(21).

$$c_1 \text{cn}(c_2, m) = A_0 \text{ and } -\sqrt{\eta' + c_1^2 \gamma}\, \text{sn}(c_2, m) \text{dn}(c_2, m) = A_{z_0}, \qquad (21)$$

through which it has been considered $A(0) = A_0$ and $m = \sqrt{\frac{c_1^2 \gamma}{2(\eta' + c_1^2 \gamma)}}$.

Assuming $\gamma = -\frac{2\eta'}{A_0^2}$ - which can be physically valid considering a large enough initial amplitude $A_0$ and huge nonlinearity $\gamma$ - and subsequently $m \to 0$, a simple form of the solution can be obtained as given in Eq.(22).

$$A(y) = A_0 \text{sech}\left(\sqrt{\eta'}\, y\right). \qquad (22)$$

This the stable solution of the nonlinear amplitude equation.

***B- Time evolution***

The necessity of investigating the temporal evolution originates from the fact that SPPs' propagation is in the form of the modulated time-dependent envelops within a waveguide. Taylor expansion of the nonlinear dispersion relation $\omega = \omega(k, A_0)$ can be written in the form of Eq. (23).

$$\omega = \omega_0 + D(k - k_0) + \frac{1}{2}\frac{\partial^2 \omega}{\partial k^2}(k - k_0)^2 + \ldots + \frac{\partial \omega}{\partial |A_0|^2}|A_0|^2 + \ldots, \quad (23)$$

where $\omega$ denotes the angular frequency and $D = \frac{\partial \omega}{\partial k}$ measures the group velocity. Re-write Eq. (23) to give Eq.(24).

$$\Omega = \kappa D + \omega'' \kappa^2 + \gamma D |A_0|^2, \quad (24)$$

in which it is assumed $\Omega = \omega - \omega_0$, $\kappa = k - k_0$ and $\omega'' = \frac{1}{2} \partial^2 \omega / \partial k^2$. Time variations of amplitude can now be explained by Eq.(25).

$$i\frac{\partial A}{\partial t} + \frac{\omega''}{2}\frac{\partial^2 A}{\partial z^2} + \gamma D |A|^2 A + i\eta \omega A = 0. \quad (25)$$

Using the Finite Difference method, one can discretize Eq.(25) as given in Eq.(26).

$$\begin{cases} \dfrac{A_j^{n+1/2} - A_j^n}{\Delta t / 2} = -\dfrac{\Delta z}{\Delta t} \cdot \dfrac{A_j^{n+1/2} - A_{j-1}^{n+1/2}}{\Delta z} \\ \quad + \dfrac{i\omega_0''}{2} \cdot \dfrac{A_{j+1}^{n+1/2} - 2A_j^{n+1/2} + A_{j-1}^{n+1/2}}{\Delta z^2} - i\gamma D |A_j^{n+1/2}|^2 A_j^{n+1/2} - \eta A_j^{n+1/2} \\ \dfrac{A_j^{n+1} - A_j^n}{\Delta t / 2} = -\dfrac{\Delta z}{\Delta t} \cdot \dfrac{A_j^{n+1/2} - A_{j-1}^{n+1/2}}{\Delta z} \\ \quad + \dfrac{i\omega_0''}{2} \cdot \dfrac{A_{j+1}^{n+1/2} - 2A_j^{n+1/2} + A_{j-1}^{n+1/2}}{\Delta z^2} - i\gamma D |A_j^{n+1/2}|^2 A_j^{n+1/2} - \eta A_j^{n+1/2}, \end{cases} \quad (26)$$

which is a half-step numerical method known as predictor-corrector method.

## C- Modulation Instability (MI)

If a preliminary condition ($\omega'' \gamma D > 0$ known as the Lighthill condition) is fulfilled inside a nonlinear optical system, temporal amplitude equation (Eq.(25)) will possesses unstable solutions in the form of quasi-periodic states. In this case, Fourier transform to the frequency domain will reveal the appearance of sidebands located at the either sides of the major frequency. Such dynamical behavior is called temporal MI[54]. Perturbations belong to the quasi-periodic state will disappear after a transient time. Subsequently, the system dynamics will return back to the regular periodic state. However, this convective feature of MI may be transformed to an absolute one if the criteria given in Eq. (27) is satisfied.

$$\gamma D A_0^2 > \frac{D^2}{4\omega''} + \omega, \quad (27)$$

which explicates that the multiplication of the initial amplitude with nonlinearity should be larger than a threshold value[54]. Through the absolute MI process, the quasi-periodic fluctuations will not damp. Instead, one can observe in frequency domain that the sidebands will grow in both number and amplitude. According to Eq.(27), those fluctuations taking the wavenumber in the range $k^2 < 4\gamma D |A_0|^2 / \omega''$ will become unstable. Hence, absolute MI can be physically interpreted as an internal feedback mechanism induced by a counter-propagating set of perturbations with negative group velocity[54].

**3- Results and discussion**

*A- Stable solution*

The effect of the nonlinearity on the SPPs can be simply estimated by the soliton norm given in Eq.(22). Fig.2 shows the results for three different values of Fermi energy $E_F$. For the considered wavelength of $\lambda = 1\,\mu m$, the lower Fermi energy level means the larger nonlinear refractive index (Eqs.(2)& (4)-(5)). On this base, the comparison between the three soliton norms in Fig.2 indicates that the plasmon soliton width will be considerably decreased if the nonlinearity is enhanced. This implies a more localized field.

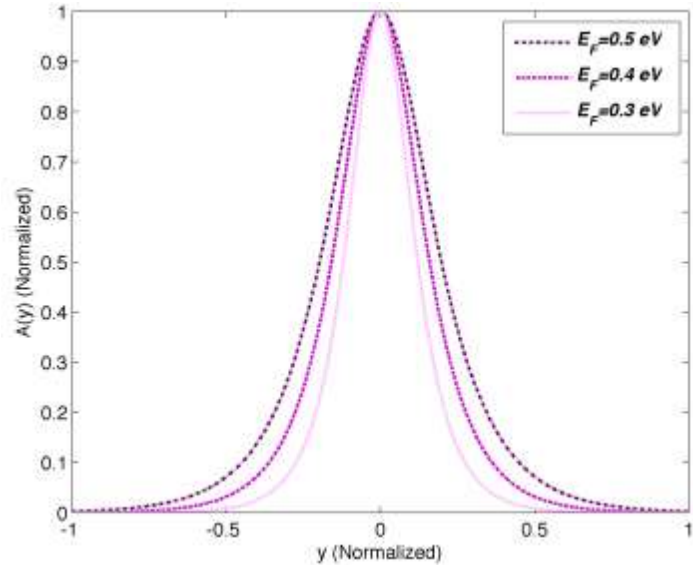

Fig.2. Comparison of widths of plasmon solitons for three different values of $E_F$. Solitons have smaller widths if propagate within a graphene sheet with lower Fermi energy level.

*B- Numerical solution, spatial MI*

As the next step, unstable solutions of Eq.(7) are to be explored when the initial amplitude of SPPs and nonlinearity are upper-threshold valued. Hence, Eq. (22) is assumed as an initial guess and Split Step Beam Propagation Method (SSBPM) is used[55]. Numerical results include transverse MI for an arbitrary propagation length (Fig.3(a) and Fig.3(b)) and variations along the propagation length (Fig.3(c) and Fig.3(d)) while two cases of $E_{F1} = 0.6\,\text{eV}$ (First column in Fig.(3)) and $E_{F2} = 0.4\,\text{eV}$ (second column in Fig.(3)) are considered. In agreement with the single layer graphene approximate lateral size, the maximum propagation length is assumed $Z = 60\,\mu m$. The two dielectric constants are $\varepsilon_1 = \varepsilon_2 = 14.5$ which can be approximately assigned to that of the ionic gel DEME-TFSI, a suitable material to provide high mobility for carriers. For simplicity, any phonon coupling effect occurred in anisotropic dielectric media is to be ignored. For the modest nonlinearity, the diffraction term in Eq.(7) surpasses and stochastic fluctuations appear(Fig.3(a) & Fig.3(c)). In contrast, for the larger nonlinearity, the competition leads to the generation of discrete plasmon solitons which in turn indicates the compact field distribution (Fig.3(b) & Fig.3(d)). In fact, for the modest nonlinearity, the transversal variations are widely distributed (Fig.3(a)) while for the larger nonlinearity, the localization appears within a compact transversal length. Consequently, a rather sensible rise in the amplitude takes place as indicated in Fig.3(b). This feature is important for designing the graphene-based electro-optical devices and cannot be resulted if the large nonlinearity regime is eliminated. With assuming the gate voltage to be $V_g = -2V$, the carrier density of $n_g = 2 \times 10^{12} \, cm^{-2}$ and the large nonlinearity regime ($E_{F1} = 0.4 \, eV$), the electric field can reach up to $\sim 50 \, kv/cm$. In comparison to the thin metal films, graphene thus supports the plasmonic waves with more intense field localization due to the larger effective area and nonlinearity. On this base, occurrence of MI with higher gain can be also anticipated.

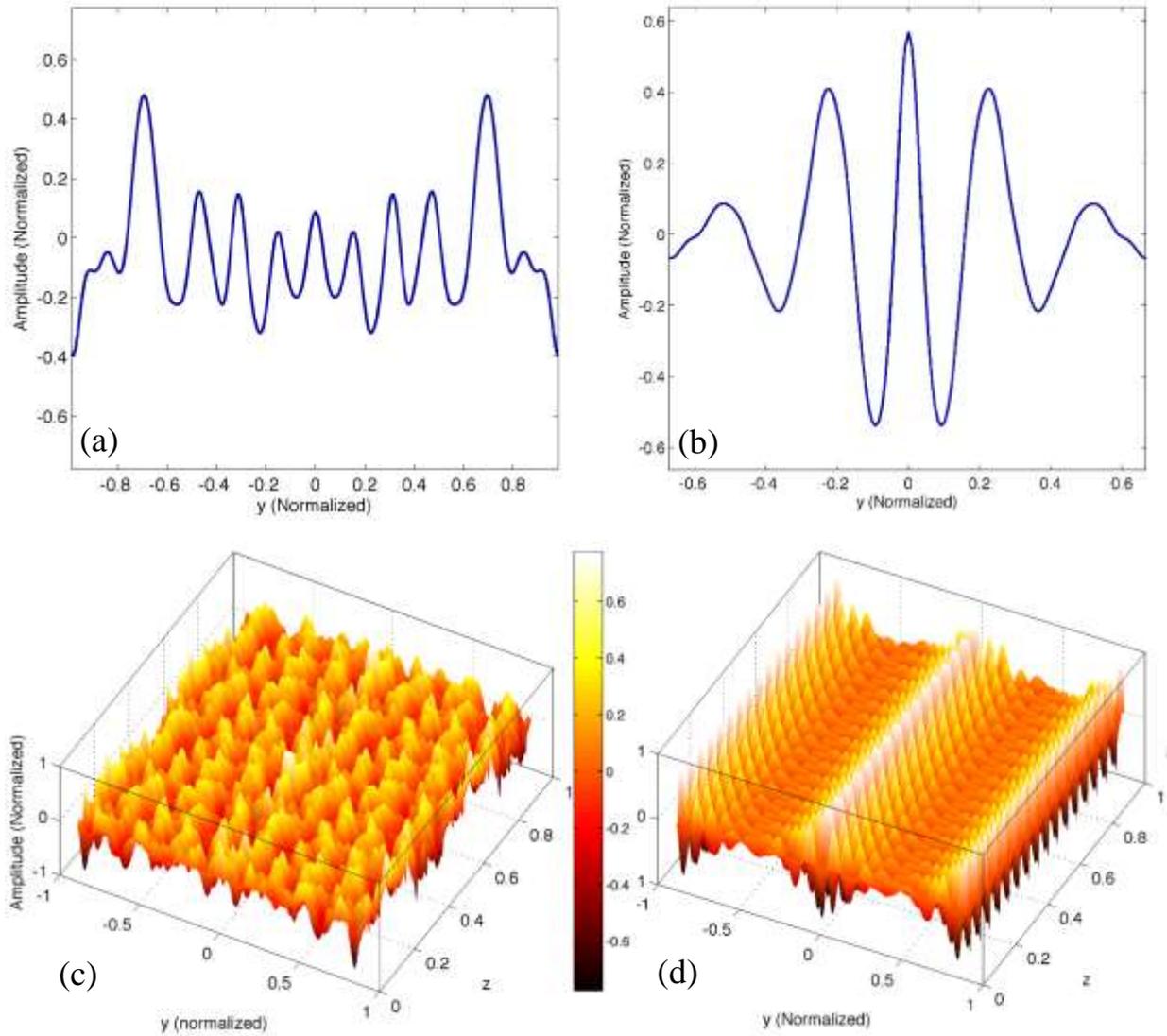

Fig.3. (a) Transversal distribution of induced plasmon solitary waves and (c) propagation along the z distance for $E_F$ = 0.6 eV, (b) & (d) the same for $E_F$ = 0.4 eV; the field is more localized for graphene sheet with lower Fermi energy level

## C- Numerical solution, temporal MI

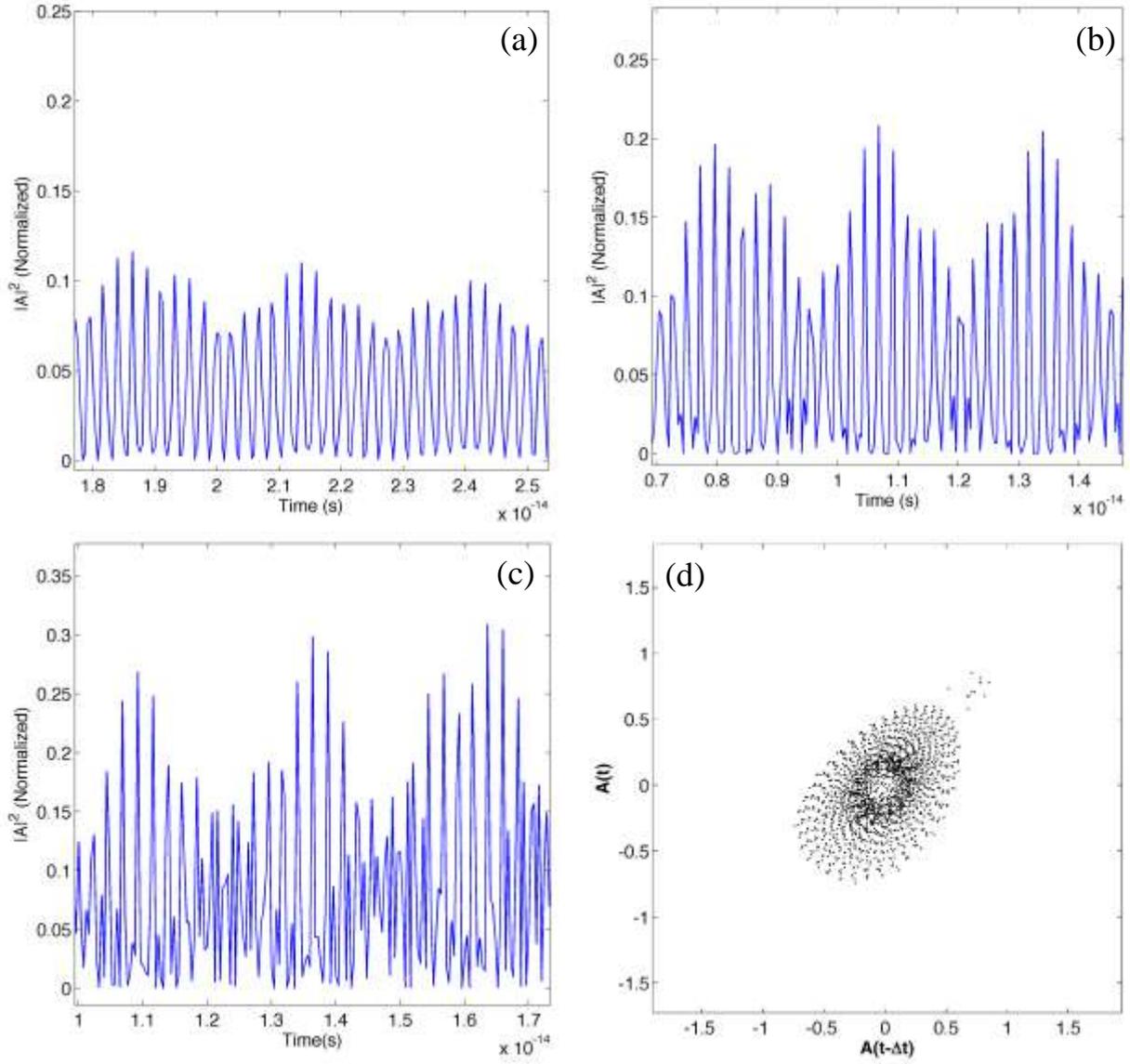

Fig.4. Multi-periodic state inferred as temporal MI obtained for (a) $\varsigma_{in} = 0.57$ and $E_F = 0.35$ eV, (b) $\varsigma_{in} = 0.75$, $E_F = 0.35$ eV, (c) $\varsigma_{in} = 0.73$ and $E_F = 0.31$ eV; (d) chaotic attractor of Fig.4(c); the character of MI will be changed from convective (Fig.4(a)) to absolute (Fig.4(c)).

In order to obtain the temporal evolution of the amplitude, Eq.(25) is numerically solved in accordance with the finite difference predictor-corrector method[56]. If the Lighthill condition is satisfied, dispersion interaction with the nonlinearity will breed the temporal MI. However, this urges a strong coupling between the external light and plasmonic waves. Accordingly, a quantity $\varsigma_{in}$ is introduced in numerical calculations to determine the coupling depth. This quantity lies between 0 and 1. Consequently, the influence of coupling depth as well as nonlinearity should be evaluated. Fig.4 shows the numerical results for three considered cases. SPPs' time evolution appears as ultrashort pulse trains. Nevertheless, for the lower coupling depth, convective MI procedure is dominant and the amplitude of plasmonic envelops damps after a transient time (Fig.4(a); $\varsigma_{in} = 0.57$, $E_F = 0.35\,\text{eV}$). For a higher coupling depth, the character of MI begins to change from the convective to absolute (Fig.4(b); $\varsigma_{in} = 0.75$, $E_F = 0.35\,\text{eV}$). Yet, the amplitude will grow fast if a larger nonlinearity is assumed (Fig.4(c); $\varsigma_{in} = 0.73$, $E_F = 0.31\,\text{eV}$). Here, the multi-frequency behavior of the ultrashort pulse trains represents a rigorous intensification. In addition, the fluctuations become complex and unpredictable. The procedure is then inferred as a route to chaos. Fig.4(d) shows the corresponding chaotic attractor. To be concise, the presumed graphene-dielectric heterostructure shown in Fig.1 is a spasing system in resemblance to a laser cavity through which QDs' pairs - formed within the surface of graphene sheet as aforementioned- are replaced with the cavity mirrors and SPPs are substituted with photons. Moreover, the multiple strongly coupled QDs' pairs will treat in analogy to the mechanism that leads to the stimulated emission for the large nonlinearity/high coupling depth if one particularly considers that the nonlinear response can be enhanced up to several orders for certain modes[13-14]. Although, the feedback mechanism is intrinsic here due to the negative group velocity assigned to the SPPs' dynamics in terahertz frequency range. If a temporal delay – explainable with quantum coherence theory- hinders the energy transfer process between each pair of QDs, fluctuations ascribed to the absolute MI may undergo a transition to chaos; conformed again to the chaotic dynamics inside a laser cavity with high power regime.

Finally, Fig. 5(a) shows SPPs' envelops in terms of Fermi energy. Obviously, the larger nonlinearity leads to the stronger field localization. Furthermore, Fig.5(b) indicates the maximum scaled values of $|A|^2$ vs. Fermi energy for different values of coupling depths. As it is clear, a rather high coupling depth will be required for realizing a large field enhancement even if the large nonlinearity regime is presumed.

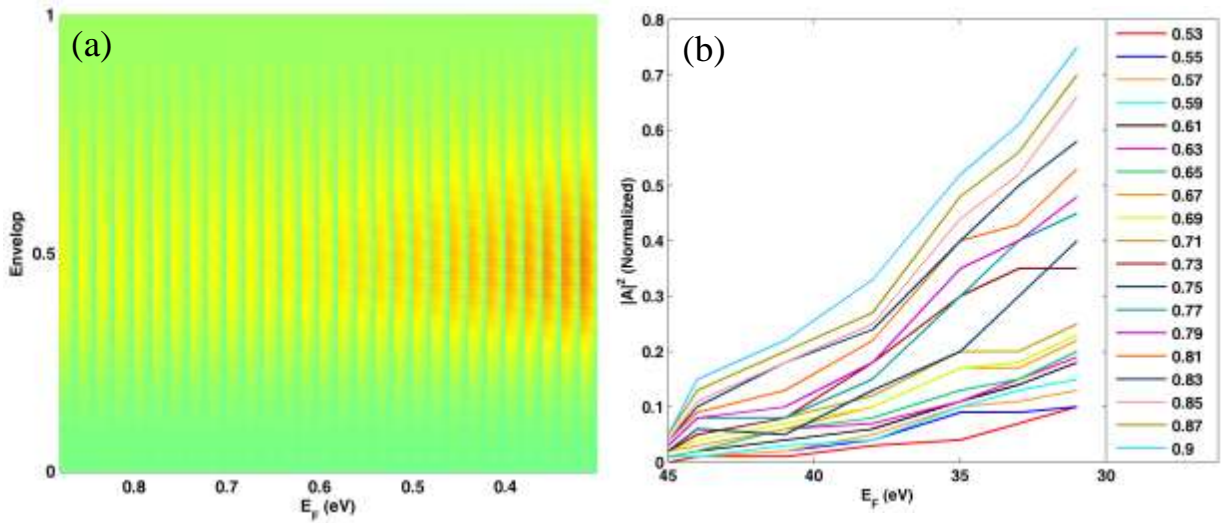

Fig.5. (a) Field envelop of SPPs vs. Fermi energy; lower Fermi level leads to more intense localization even for temporal envelops, (b) Maximum amount measured for $|A|^2$ vs. Fermi energy for different values of coupling depths; growth rate is higher for the higher coupling depths.

## 4- Conclusion

An analytical solution of nonlinear plasmon amplitude equation based on JEFs has shown that the large nonlinearity leads to the compact field distribution for a plasmon soliton propagating within a surface of single layer graphene embedded inside two dielectric media. Then, a numerical solution of the amplitude equation using SSBPM has revealed that the discrete plasmon solitons can be only generated if the large nonlinearity regime is assumed. On the side, if the temporal variations are also taken into account, ultrashort pulse trains with multi-periodic fluctuations inferred as temporal MI can be resulted. However, the character of MI can be altered from convective to absolute if the coupling depth between the QDs formed within the surface of graphene sheet is raised enough. This coupling has an intrinsic nature. Consequently, the amplitude of SPPs can grow in such a manner that the multi-periodic behavior is unpredictable, a transition to chaos.

Through this study, the numerical calculations have been implemented in the vicinity of critical cut-off frequency beyond which the complex chaotic regime is deduced. Nonetheless, this brings difficulties about the possibility of appearing infinite answers. If a general amplitude equation is alternatively used, the results can give more details about the MI behavior. Furthermore, a graphene-based waveguide can be considered with more than one suspended graphene sheet. This will induce a significant enhancement of the nonlinearity and thus, can considerably decrease the power consumption for electro-optical applications in which the nonlinear response and subsequently, the chemical potential is to be controlled by an applied bias voltage. On the other hand, since the graphene is a biocompatible nanomaterial, the garphene-based SPP waveguide can be used for biomedical diagnostic application[57].